\documentclass[twocolumn,amsmath,amssymb]{revtex4-2}
\usepackage{graphicx,hyperref}
\usepackage{color}
\usepackage{caption}
\usepackage{subcaption}
\usepackage{dcolumn}

\begin{document}

\title{Nuclear shape / phase transitions in the N = 40, 60, 90 \\ regions}

\author
{Dimitrios Petrellis$^1$, Adam Pr\'{a}\v{s}ek$^{2,\dagger}$, Petr Alexa$^2$, Dennis Bonatsos$^3$, Gabriela Thiamov\'{a}$^4$, and Petr Vesel\'{y}$^1$}

\affiliation
{$^1$Nuclear Physics Institute, Czech Academy of Sciences, CZ-250 68 \v{R}e\v{z} near Prague, Czech Republic}

\affiliation
{$^2$Department of Physics, V\v{S}B -- Technical University Ostrava,
17. listopadu 2172/15, CZ-708 00 Ostrava, Czech Republic}

\affiliation
{$^3$Institute of Nuclear and Particle Physics, National Centre for Scientific Research ``Demokritos'', GR-15310 Aghia Paraskevi, Attiki, Greece}

\affiliation
{$^4$Universite Grenoble 1, CNRS, LPSC, Institut Polytechnique de  Grenoble, IN2P3, F-38026 Grenoble, France}

\affiliation
{$^\dagger$deceased}

\begin{abstract}
We investigate the isotopes of Se, Zr, Mo and Nd in the regions with N = 40, 60 and 90, where a first-order shape / phase transition, from spherical to deformed,  can be observed. The signs of phase transitional behavior become evident by examining structure indicators, such as certain energy ratios and B(E2) transition rates and, in particular, how they evolve with neutron number. Microscopic mean-field calculations using the Skyrme-Hartree-Fock + Bardeen-Cooper-Schrieffer framework also reveal structural changes when considering the evolution of the resulting potential energy curves as functions of deformation. Finally,  macroscopic calculations, using the Algebraic Collective Model, specifically for $^{74}$Se, $^{102}$Mo and $^{150}$Nd, after fitting its parameters to experimental spectra, result in potentials that resemble some of the potentials proposed in the framework of the Bohr Hamiltonian to describe shape transitions in nuclei. A more detailed account can be found in \cite{Prasek2024}.
\end{abstract}

\maketitle

\section{Introduction}
\label{intro}
Changes in the shapes of atomic nuclei have been the subject of intensive study for several decades \cite{Casten2009, Cejnar2009, Cejnar2010}. Already in 1977 a mechanism based on the shell-model was proposed to explain the sudden onset of deformation for certain neutron numbers in even-even nuclei \cite{Federman1977}. A few years later, the Interacting Boson Model \cite{Iachello1987} provided a formal framework, through its classical limit, for the description of such phenomena as transitions between its symmetries, giving birth to the concept of shape / phase transitions (SPT). Specifically, a second-order SPT was identified between the U(5) (spherical) and O(6) ($\gamma$-unstable) limit and a first-order SPT between the U(5) and the SU(3) (axially deformed) limit  \cite{Dieperink1980, Feng1981}. Twenty years later, the E(5) and X(5) critical point symmetries (CPS) were introduced, as equivalent descriptions of these SPTs, within the framework of the Bohr Hamiltonian \cite{Iachello2000, Iachello2001}. In the present work, we investigate the occurrence of the first-order SPT  in the regions with neutron numbers $N = 40,~60,~90$, both from a microscopic, as well as macroscopic point of view.

\section{Manifestations of structural change}
\label{sec-1}
The importance of the $N=60$ and $N=90$ regions for the appearance of deformation was already highlighted in \cite{Federman1977}. A recent study \cite{JPG2023} indicated a structural similarity between the $N=40$ region around $^{74}$Se and the $N=60,~90$ regions, based on systematics of energy levels and $B(E2)$ transition rates. Therefore, a common theoretical framework for these regions is deemed necessary. 
\begin{figure*}[]
      \centering
	   \begin{subfigure}{0.27\linewidth}
		\includegraphics[width=47mm]{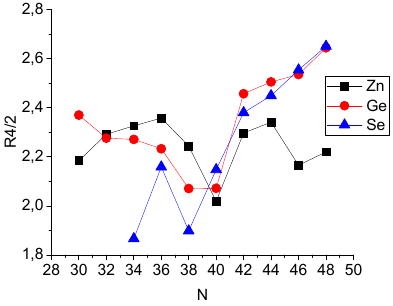}
		\label{fig:R42_1}
	   \end{subfigure}
	   \begin{subfigure}{0.25\linewidth}
		\includegraphics[width=47mm]{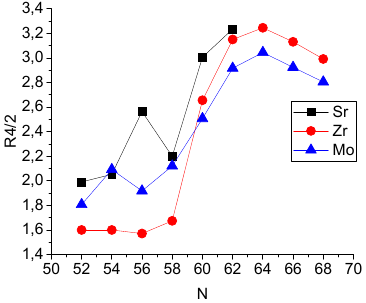}
		\label{fig:R42_2}
	   \end{subfigure}
	   \begin{subfigure}{0.26\linewidth}
		\includegraphics[width=47mm]{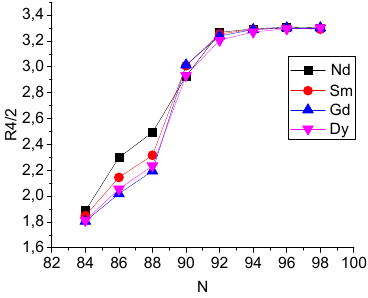}
		\label{fig:R42_3}
	   \end{subfigure}
	\vfill
	   \begin{subfigure}{0.27\linewidth}
		\includegraphics[width=47mm]{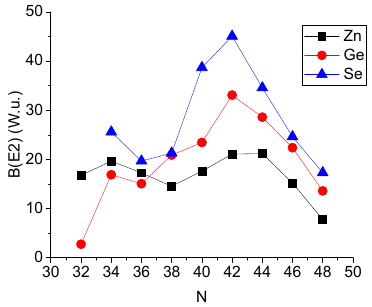}
		\label{fig:BE2_1}
	   \end{subfigure}
	   \begin{subfigure}{0.26\linewidth}
		\includegraphics[width=47mm]{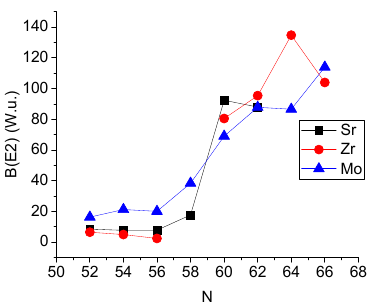}
		\label{fig:BE2_2}
	   \end{subfigure}
	   \begin{subfigure}{0.26\linewidth}
		\includegraphics[width=47mm]{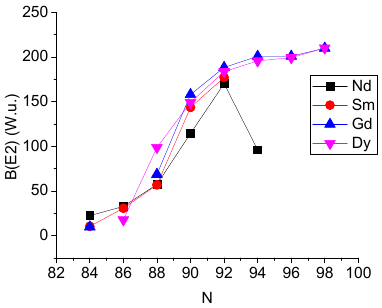}
		\label{fig:BE2_3}
	   \end{subfigure}
	\vfill
	   \begin{subfigure}{0.27\linewidth}
		\includegraphics[width=47mm]{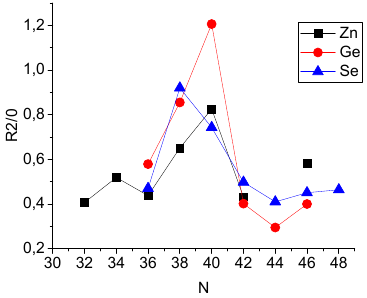}
		\label{fig:R20_1}
	   \end{subfigure}
	   \begin{subfigure}{0.26\linewidth}
		\includegraphics[width=47mm]{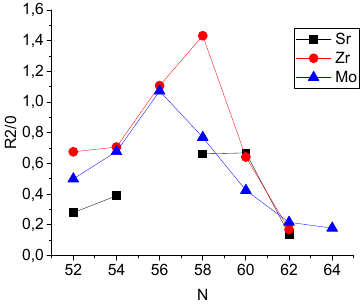}
		\label{fig:R20_2}
	   \end{subfigure}
	   \begin{subfigure}{0.26\linewidth}
		\includegraphics[width=47mm]{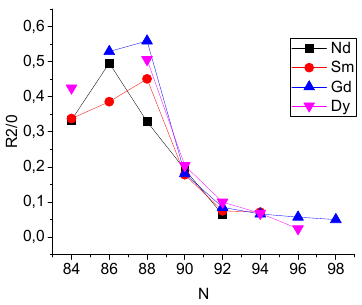}
		\label{fig:R20_3}
	   \end{subfigure}
	\vfill
	   \begin{subfigure}{0.27\linewidth}
		\includegraphics[width=47mm]{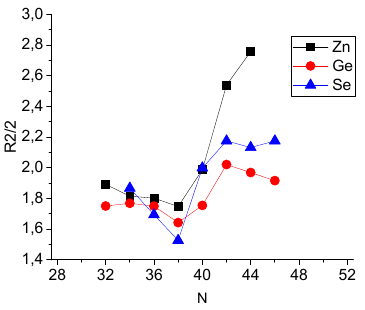}
		\label{fig:R22_1}
	   \end{subfigure}
	   \begin{subfigure}{0.26\linewidth}
		\includegraphics[width=47mm]{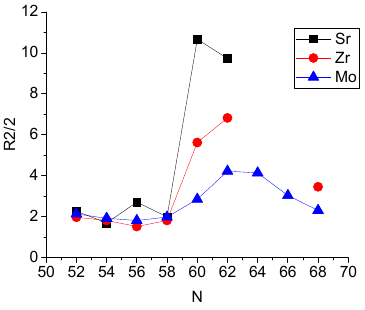}
		\label{fig:R22_2}
	   \end{subfigure}
	   \begin{subfigure}{0.26\linewidth}
		\includegraphics[width=47mm]{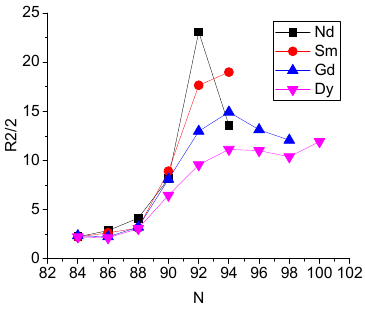}
		\label{fig:R22_3}
	   \end{subfigure}
	\caption{Experimental energy ratios $R_{4/2}$, $R_{2/0}$, $R_{2/2}$ and $B(E2; 2_1^+ \to 0_1^+)$ transition rates as functions of neutron number $N$, in the $N=40,~60,~90$ regions. Adapted from \cite{Prasek2024}.}
	\label{fig:Ratios}
\end{figure*}

In  Fig. \ref{fig:Ratios} we present experimental \cite{ensdf} energy ratios defined as
\begin{equation}
R_{4/2} = \frac{E(4_1^+)}{E(2_1^+)}, \quad R_{2/0} = \frac{E(2_1^+)}{E(0_2^+)}, \quad R_{2/2} = \frac{E(2_\gamma^+)} {E(2_1^+)},
\end{equation}
as well as the experimental $B(E2; 2_1^+ \to 0_1^+)$ transition rates \cite{Pritychenko} for nuclei in all three regions. The nuclei under consideration are Zn-Ge-Se for the $N=40$ region, Sr-Zr-Mo for the $N=60$ region and Nd-Sm-Gd-Dy for the $N=90$ region.
The ratio $R_{4/2}$ is a well known indicator of collectivity \cite{Casten1990},while the transition rate $B(E2; 2_1^+ \to 0_1^+)$ is known to be proportional to the square of the quadrupole deformation parameter $\beta$. The ratio $R_{2/0}$ is known \cite{JPG2023} to exhibit a maximum in the region of the SPT from spherical to deformed nuclei. Apart from the obvious sudden changes, when crossing the specified neutron numbers, all the observed trends in Fig. \ref{fig:Ratios} are compatible with a relative drop in the $2_1^+$ level, which is a signature of increased collectivity and subsequent deformation \cite{Casten1990}. It should be noted that the $R_{4/2}=2.904$ and $R_{2/0}= 0.177$ values characteristic of the X(5) predictions \cite{Iachello2001,Bonatsos2004} apply sufficiently well only in the case of the $N=90$ isotones \cite{Krucken2002,Casten2001,Tonev2004,Caprio2002}. The values observed in the $N=40$ region, in particular, are close to the $R_{4/2}=2.44$ and $R_{2/0}= 0.35$ values predicted by the X(3) model \cite{Bonatsos2006}, which is a $\gamma$-rigid version of X(5).

\section{Microscopic self-consistent calculations}
\label{sec-2}
The evolution of potential energy curves (PECs) as functions of deformation, with  increasing number of valence nucleons offers one of the best ways to illustrate ground state phase transitions \cite{Casten2009, Cejnar2009, Cejnar2010}. In this section we present the results of Skyrme-Hartree-Fock + Bardeen-Cooper- Schrieffer (SHF + BCS) calculations \cite{Bender2003} performed with the code SkyAx \cite{Reinhard2021}, which uses Nilsson orbitals as initial single particle wave functions and is therefore suitable for axially deformed nuclei.
\begin{figure*}[]
      \centering
	   \begin{subfigure}{0.46\linewidth}
		\includegraphics[width=\linewidth]{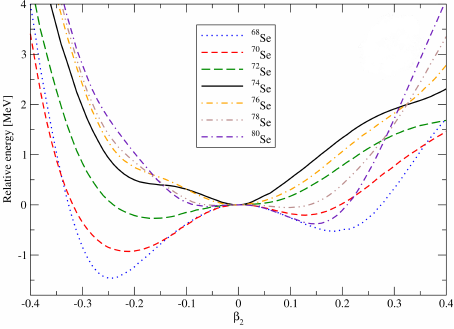}
		\label{fig:Se_pec1}
	   \end{subfigure}
	   \begin{subfigure}{0.46\linewidth}
		\includegraphics[width=\linewidth]{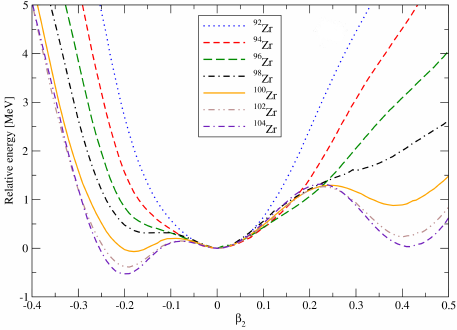}
		\label{fig:Zr_pec12}
	    \end{subfigure}
	\vfill
	     \begin{subfigure}{0.46\linewidth}
		 \includegraphics[width=\linewidth]{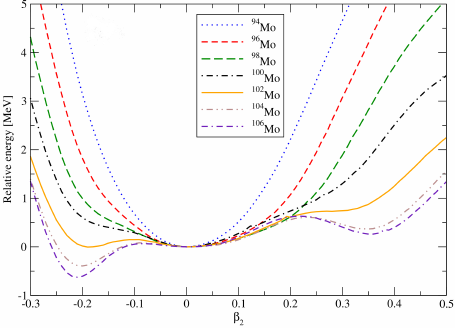}
		 \label{fig:Mo_pec1}
	      \end{subfigure}
	       \begin{subfigure}{0.46\linewidth}
		  \includegraphics[width=\linewidth]{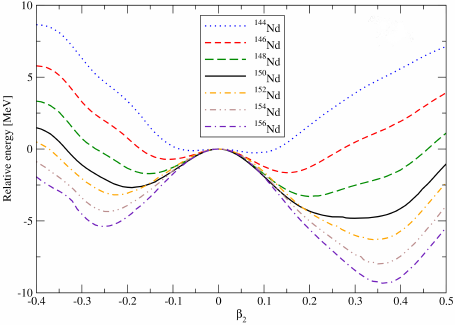}
		  \label{fig:Nd_pec1}
	       \end{subfigure}
	\caption{PECs for Se, Zr, Mo, and Nd isotopes, as functions of quadrupole deformation, resulting from SHF+BCS calculations with the SV-bas parametrization. Adapted from \cite{Prasek2024}.}
	\label{fig:PECs}
\end{figure*}
By imposing a quadrupole constraint these calculations produce PECs, as functions of the quadrupole deformation $\beta_2$. Fig. \ref{fig:PECs} depicts the PECs that resulted from the SV-bas parametrization  \cite{Klupfel2009}. As can be seen, the Se isotopes exhibit a transition from an oblate deformation ($\beta_2 <0$) at $N=34$ to a prolate one ($\beta_2>0$) at $N=46$, after passing through a spherical one at $N=40$. In the case of Zr and Mo isotopes, a spherical shape at $N=52$ evolves into a deep oblate minimum and two shallower ones, a spherical and a prolate one at $N=64$. At $N=60$ the oblate and the spherical minima are close in energy while the prolate one appears higher. Finally, in the case of Nd isotopes a relatively flat PEC at $N=84$ evolves at $N=90$ into one with two minima, one prolate and one oblate, that are almost degenerate, while at $N=96$ the prolate minimum has prevailed. The phase coexistence characteristic of first-order phase transitions is present in all cases.
\section{Algebraic Collective Model calculations}
\label{sec-3}
\begin{figure*}[]
\centering
\vspace*{1cm}
\includegraphics[width=8cm,clip]{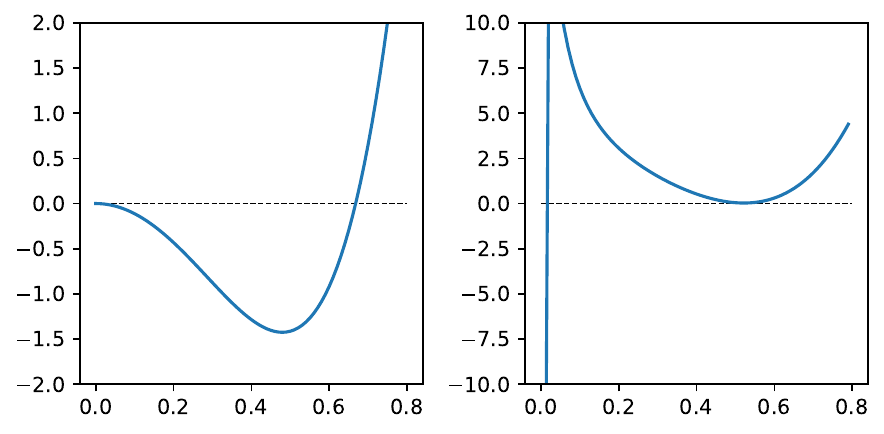}
\caption{Part of the ACM potential containing only positive powers of $\beta$ (left) and the total ACM potential (right), obtained from the fit to the experimental spectra of $^{150}$Nd. Adapted from \cite{Prasek2024}.}
\label{fig:ACM_150Nd}
\end{figure*}

The algebraic collective model (ACM) \cite{Rowe2009,Rowe2010} is a computationally tractable version of the collective model of Bohr and Mottelson. 
\begin{figure*}[]
      \centering
	   \begin{subfigure}{0.46\linewidth}
		\includegraphics[width=\linewidth]{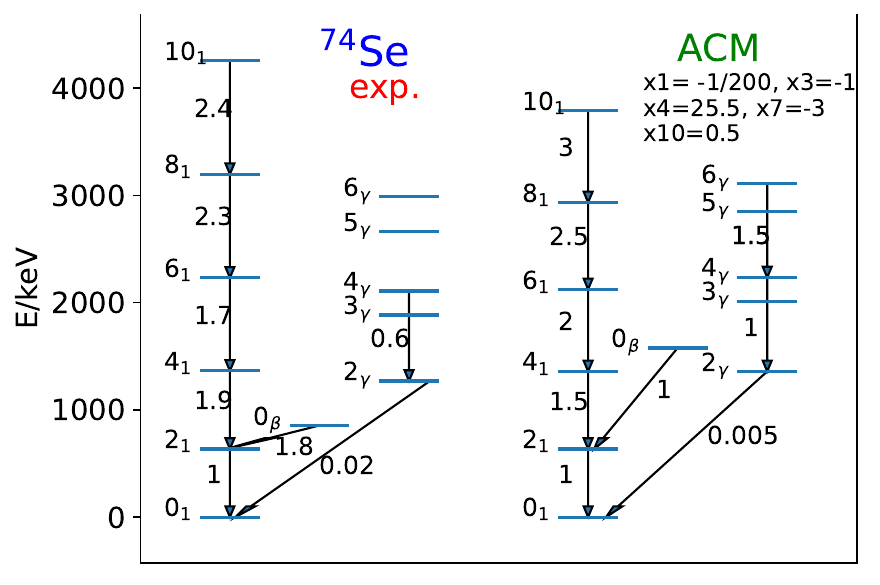}
		\label{fig:74Se_enls}
	   \end{subfigure}
	   \begin{subfigure}{0.46\linewidth}
		\includegraphics[width=\linewidth]{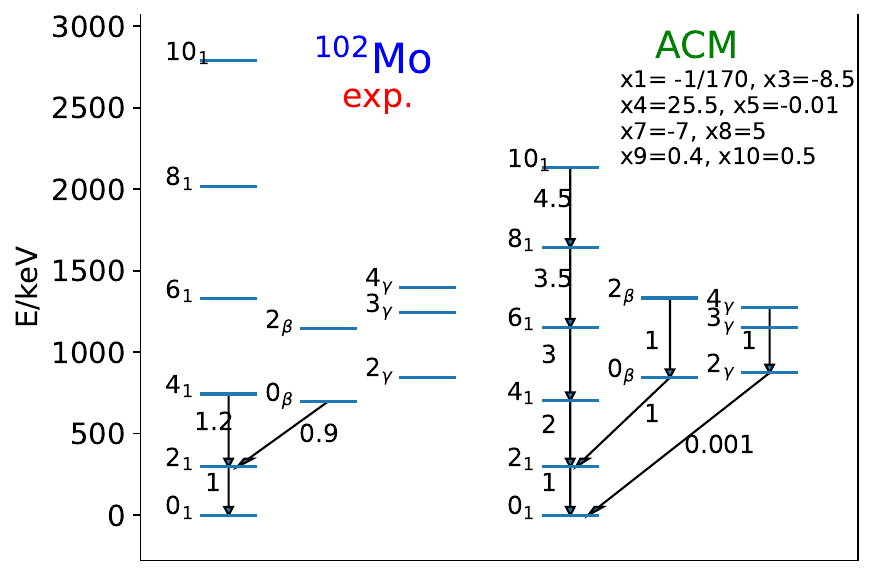}
		\label{fig:102Mo_enls}
	   \end{subfigure}
	  \vfill 
	   \begin{subfigure}{0.46\linewidth}
		\includegraphics[width=\linewidth]{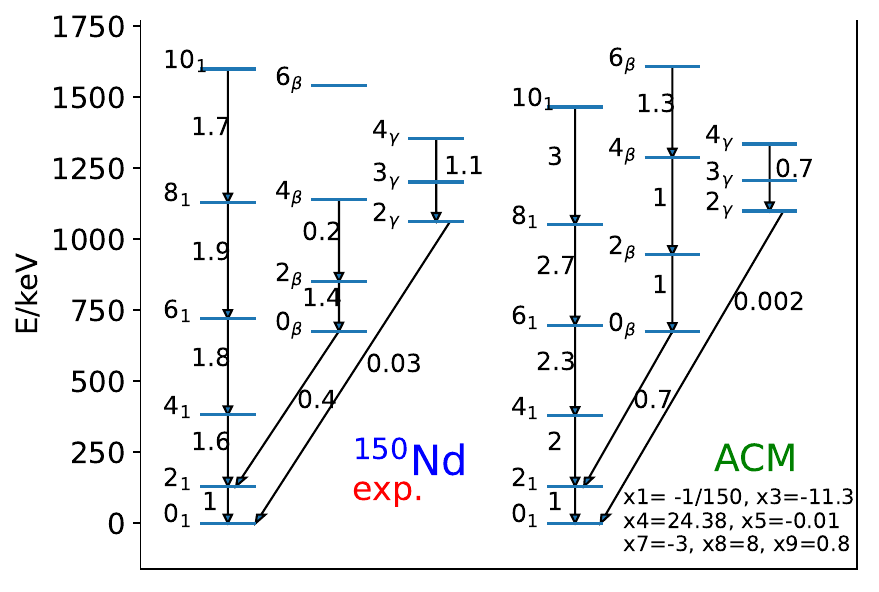}
		\label{fig:150Nd_enls}
	   \end{subfigure}
	\caption{Comparison of experimental (left) and ACM (right) spectra for $^{74}$Se, $^{102}$Mo and $^{150}$Nd. The energy differences of the $2^+_1$ and $0^+_1$ states are normalized 
to the experimental values. The $B(E2)$ transition rates are expressed in units of the $B(E2,2_{1}^{+}\rightarrow 0_{1}^{+}) = 1$ transition. The parameters used in the ACM calculations are also displayed. Adapted from \cite{Prasek2024}.}
	\label{fig:ENLs}
\end{figure*}
Matrix elements can be analytically calculated by exploiting the model's algebraic structure, which is that of the ${\rm SU}(1,1) \times {\rm SO}(5)$ dynamical group. An  appropriate basis of wave functions is used to diagonalize the following Hamiltonian  
\begin{eqnarray}
\hat H &=& x_{1}{\nabla^2}+x_{2}+x_{3}\beta^{2}+x_{4}\beta^{4}+\frac{x_{5}}{\beta^{2}}+x_{6}\beta \cos{3\gamma} \nonumber \\
&+&x_{7}\beta^3 \cos{3\gamma}+x_{8}\beta^5 \cos{3\gamma} + \frac{x_{9}}{\beta}\cos{3\gamma} \nonumber \\ 
&+&x_{10}\cos^2{3\gamma}+x_{11}\beta^{2}\cos^{2}{3\gamma}+x_{12}\beta^{4}\cos^{2}{3\gamma} \nonumber \\ \nonumber \\
&+&\frac{x_{13}}{\beta^{2}}\cos^{2}{3\gamma}+\frac{x_{14}}{\hbar^{2}}[\hat \pi \otimes \hat q\otimes \hat \pi]_0, 
\label{HACM}
\end{eqnarray}
where
\begin{equation}
 \nabla^2 = \frac{1}{\beta^4} \frac{\partial}{\partial\beta} \beta^4 \frac{\partial}{\partial\beta} + \frac{1}{\beta^2} \hat\Lambda
\end{equation}
is the  Laplacian on the 5-dimensional collective model space  and $\hat\Lambda$ is the SO(5) angular momentum operator. A computer code for performing such calculations is available \cite{Welsh2016}, while the $x_1,\dots,x_{14}$ parameters are fitted to data. Numerical results for $^{74}$Se, $^{102}$Mo and $^{150}$Nd are shown in Fig. \ref{fig:ENLs}. One can make some qualitative observations by looking at the various parts of the potential. For example, the potentials that result from the $^{150}$Nd fit are plotted in Fig. \ref{fig:ACM_150Nd}. The part containing only positive powers of $\beta$ looks like a Davidson potential, widely used in the context of nuclear shape phase transitions \cite{Bonatsos2007}, although without its divergent behavior at the origin. The total potential resembles the so called confined beta-soft (CBS) model \cite{Pietralla2004}, which uses an infinite square well potential in $\beta$ displaced from the origin and which is applicable both in deformed and transitional nuclei.
\section{Summary and outlook}
\label{sec-4}
In the present work we studied the $N = 40,~60,~90$ regions of the nuclear chart by considering empirical systematics along with microscopic mean-field and macroscopic collective model calculations. The systematics of energy spectra and $B(E2)$ transition rates reveal the existence of a first-order QPT in all three cases, although only in the $N = 90$ case deformed values are achieved, corresponding to the X(5) CPS results. The $N = 40,~60$ regions eventually lead to more moderate deformations with results closer to those of the X(3) CPS. The microscopic calculations, of the Skyrme-Hartree-Fock + BCS type, and the resulting PECs reveal a tendency towards prolate shapes in the $N = 90$ region and towards oblate shapes in the $N = 40, ~60$ regions. The macroscopic ACM calculations, also produce PECs that resemble potentials that were used in the framework of the Bohr Hamiltonian to describe nuclear SPTs.

Some points require further research. A possible correspondence between the PECs resulting from the microscopic calculations and those resulting from the ACM could provide a useful link between the microscopic and macroscopic pictures. Another question to be addressed is the preference for prolate shapes in the $N = 90$ region and for oblate shapes in the $N = 40, ~60$ regions. Finally, it should be mentioned that the $N = 40$ region is special, in that the valence  protons and neutrons occupy the same major shell. Thus, another mechanism than the one proposed in \cite{Federman1977} may be required to explain shape / phase transitional behavior in that region.

\section{Acknowledgements}
\label{sec-5}

This work was supported by the project SP2023/25 financed by the Czech Ministry of Education, Youth 
and Sports and by the Czech Science Foundation grant GA22-14497S. 
This paper is dedicated to the memory of Adam Pr\'{a}\v{s}ek, an excellent student of V\v{S}B-Technical University of Ostrava who worked on the ACM calculations
and passed away in Grenoble Alpes on October 1, 2023.

\end{document}